\def\bi{\bigskip}
\def\be{\begin{equation}}
\def\en{\end{equation}}
\def\bq{\begin{eqnarray}}
\def\eq{\end{eqnarray}}
\def\noi{\noindent}

\documentstyle[12pt,epsf]{article}

\begin{document}
\begin{titlepage}
%\begin{flushright}
%hep-ph/0311133\\
%Nov. 25 2002
%version 1
%\end{flushright}
\vspace*{0.6cm}

\begin{center}
{\Large\bf $\pi\pi$ Invariant mass spectrum in $V'\to V~\pi\pi$ \\ \vspace*{0.3cm}
and the $f_0(600)$ pole.}
%{\footnotesize\it $^1$Departament de F\'{\i}sica,}

\vspace*{0.8cm}

{\large A. Gallegos$^1$, J.L.~Lucio M$^1$ and J. Pestieau $^2$}\\

\vspace*{0.8cm}

$^1$ Instituto de F\'{\i}sica, Universidad de Guanajuato,
Lomas del Bosque {\#}103, Lomas del Campestre, 37150 Le\'on, 
Guanajuato, Mexico\\
\vspace*{0.4cm}

$^2$ Institut de Physique Th\'eorique, Universit\'e  Catholique de Louvain\\
Chemin du Cyclotron 2, B-1348 Louvain la Neuve, Belgium\\

\end{center}
\vspace*{5.50cm}

%\begin{abstract}
\noi  We consider the phenomenological description of the two pion invariant 
mass spectrum in the $V' \to V\pi\pi$ decays. We study the parametrization of 
the amplitude involving both $S$ and $D$ wave contributions. From a fit to the
two pion decays of the $\Upsilon(nS)$ and $\Psi(nS)$ we determine the 
$f_0(600)$ mass and width to be $m_{f_0}=528 \pm 32$ MeV and $\Gamma_{f_0}=413
 \pm 45$ MeV. The mass and width values we report correspond  to the real 
and imaginary part of the S matrix pole respectively.

%\end{abstract}
\vspace*{1.5cm}

\end{titlepage}
%\section{Introduction}
\newpage

 The experimental identification of low mass scalar resonances is a long 
standing puzzle whose origin can be traced back to some of the following 
characteristics: a large decay width, possible mixing with multiquark or glue 
balls, overlap of resonances and the opening of channels, responsible for the 
appearance of cusp effects \cite{pdg}. In particular the $f_0(600)$ has a long
 history, it has been included in some issues of the Review of Particle 
Properties but it has also been excluded for long periods by the Particle Data
Group. Recently, experimental evidence from different corners of particle 
physics  has accumulated 
confirming the existence of the $f_0(600)$ resonance. There have been attempts
to interpret the low lying scalars as multiquark states \cite{jaffe,close}
or $\bar KK$ bound states \cite{fazio,dean}. Also, models exist starting 
directly from chiral symmetry \cite{scadron,torn,mauro} or else unitarized 
models where the scalar nonet arises \cite{oller,beveren}. However the 
understanding of actual processes involving scalar mesons for descriptions 
starting from first principles are not available, effective theories are not 
well suited
to deal with these scalars, the use of sum rules is perhaps the best approach
to the problem\cite{golkap}, however no predictions for all of the scalars 
have been 
advanced in that framework. Furthermore, the phenomenological description of 
broad resonance faces severe problems, the determination of the physical 
parameters -mass and width- is a non trivial problem that requires  study of 
the non-resonant background dependence.
\bi

For small invariant mass of the pion pair, data 
is dominated by the phase shift \cite{Roos,E865}. Studies have been 
carried in this kinematical region using chiral symmetry and Roy equations
-solid theoretical tools- establishing thus a firm result to be considered
by other analysis \cite{Cola}. Inclusion of the di-pion low 
invariant mass favor a light and very broad $f_0(600)$ ($m\approx 470 \pm
30 MeV$
and $\Gamma/2\approx 295 \pm 20$ MeV). Below 1 GeV, information on the $\pi\pi$
phase shift is obtained from $\pi N$ scattering, $P\bar P$ annihilation at rest
and central production, these data allow the existence of a broad 
($\Gamma\approx 500 MeV$) scalar meson resonance \cite{Gray,Kamisnsky}. 
Decay of pseudoscalar charmed 
mesons are also a source of valuable data involving scalar mesons, although
the $f_0(600)$ has only been reported by the E-791 collaboration after the 
Dalitz plot analysis of $D \to \pi^+\pi^-\pi^+$ \cite{E791}. 
Another important source of 
information are the vector meson decays. The processes considered involve 
either the scalars themselves or pion pairs together with photons or vector 
mesons. Among these we can mention $(\rho,\omega,\Phi) \to P~\bar P~\gamma$ 
\cite{achasov,Ak,acha,Anto,KLOE}, $J/\Psi \to P~\bar P~V$, $\Psi'\to\Psi 
~P~\bar P$, and $\Upsilon(nS)\to\Upsilon(mS)P~\bar P$ \cite{Butler},
where $P$ stands for a pseudo scalar ($\pi$ or $K$) and $V$ for a vector meson
($\rho,\omega$). Experimental evidence for the contribution of scalar
resonances to some of these processes has been reported, and a recent 
analysis of the data for $\Upsilon(nS)\to\Upsilon(mS)P~\bar P$ concludes the 
contribution of the  $f_0(600)$ with a large uncertainty in the width
($m=526^{+48}_{-37}$, $\Gamma= 301 ^{+145}_{-100} MeV$) 
\cite{Komada}. It should also be noticed that
in the experimental data for $\psi(2S)\to\pi^+\pi^-J/\Psi$  reported by the 
BES collaboration \cite{bai}, no evidence for $f_0(600)$ contribution is 
foreseen.
\bi

In this paper we concentrate on the decay of heavy quark vector meson 
resonances ($\Upsilon$, and $\Psi$), where pair of pions are produced with 
invariant mass ranging 
below the 1 GeV region. The kind of processes we are interested in has been 
considered by a number of authors using techniques as diverse as pure chiral 
symmetry \cite{brown}, non-relativistic theory assuming the existence of 
an Adler zero \cite{pham}, the color field multipole expansion in the non-
relativistic limit \cite{Voloshin,Novikov}, effective Lagrangian based on
chiral symmetry and the heavy quark expansion \cite{Mannel,Yan,chen}, and also
a purely phenomenological description based on a Breit-Wigner parametrization \cite{Komada}.
The framework so developed is used then to describe the two pion invariant
mass spectrum and, in some cases, also the angular distribution. The latter is
important since existing  experimental data can discriminate among the 
proposed models. Furthermore, as far as we know, the complete 
parametrization of the amplitude has not been discussed and some confusion 
has arisen concerning the ``S'' and ``D'' wave contributions to the amplitude \cite{chen}.
\bi

Our purpose is to perform an analysis as general as possible, including both
$S$ and $D$ waves in the ($ P\bar P$) system. We show that Lorentz covariance 
fixes the parametrization
of the amplitude in terms of four form factors. Using the two gluon mechanism
implied by the OZI rule, the identification of the spin zero (di-pion S wave)
and spin two (di-pion D wave) are unambiguously identified. For simplicity we
restrict our analysis to two form factors, $a_0(m_{\pi\pi})$  and  $a_2(m_
{\pi\pi})$ which are associated to S and D waves respectively. We then invoke 
the pole approach to parametrize $a_0, a_2$. We propose a Breit-Wigner plus
a background for $a_0$ and a soft background for $a_2$. In this way we claim
that crossed channel as well as higher scalar resonance contributions are 
taken into account. Thus, we expect that the amplitude associated to crossed
channel and  higher resonance contribution behave as a soft function in the 
500-800 MeV range, where the $f_0(600)$ is expected to lie. Our strategy is to
perform a joint fit to  data from different processes, involving 165 points 
and 33 parameters. 
\bi

The set of data points we consider include the $\pi-\pi$ invariant mass 
distribution of the  $\Upsilon(3S) \to \Upsilon(1S)+\pi\pi$ \cite{Butler}, 
$\Upsilon(3S) \to \Upsilon(2S)+\pi\pi$ \cite{Butler}, $\Upsilon(2S) \to 
\Upsilon(1S)+\pi\pi$ \cite{Butler,alexander}, $\psi(2S) \to J/\psi +\pi\pi$ 
\cite{bai} decays. The following characteristics are worth-mentioning: 
{\it i}) we are considering flavor conserving processes, and all of them are 
expected to proceed through two gluons. This point will be relevant when
discussing the parametrization of the form factors. {\it ii}) the smallness 
of the phase space available for the processes under consideration ($2m_{\pi}
\leq \sqrt{s}\leq 0.9~GeV$). Note that the expected central value and the 
large width of the $f_0(600)$ would imply non negligible resonance effects 
in these processes and {\it iii}) the invariant 
mass distribution for $s \to s_{th}$ -where $s_{th}$ stands for the threshold 
value of the di-pion invariant mass- show a peculiar behavior to be 
contrasted with the typical ($s-2m^2_\pi$) expected in processes involving 
soft pions. Compare for example in Fig(1) the threshold behavior of the 
$\sqrt{s}$ distribution for $\Upsilon(2S) \to \Upsilon(1S)+\pi\pi$ or 
$\psi(2S) \to J/\psi+\pi\pi$ with $\Upsilon(3S) \to \Upsilon(1S)+\pi\pi$.
\bi

In order to understand the nature of the problem we face, remark that the more
recent data \cite{alexander,bai} have been analyzed in terms of the 
scale anomaly. Indeed, these processes can be fitted without problem, which
brings the question if the full set of data we consider can be explained using
the same formalism. Our results show that this is not the case and that 
inclusion of the $f_0(600)$ improves our understanding of the data. Thus, as 
far as we can see, any attempt to provide a successful description of the 
flavor conserving two pion decays of the $\Upsilon$ and $\psi$ families 
should consider the full set of data, since the foreseen physical mechanisms
(scale anomaly, scalar resonance exchange) are operative in all cases under
consideration.
\bi

\section{Parametrization of the amplitude}.

We are interested in the decay $V'(p', \eta')\to V(p,\eta)\pi(p_1)\pi(p_2)$
where the letters in parenthesis stand for the four-momenta and polarization
of the corresponding particles. We introduce $q\equiv p_1-p_2$ and $Q\equiv
p_1+p_2$. $Q^2=s$, and $p'^2=m'^2, p^2=m^2, {p_1}^2={p_2}^2 =m^2_\pi.$ In 
order to obtain the general parametrization is convenient to consider the 
amplitude for the exchange of arbitrary spin zero and spin two meson like 
objects (although we do not consider the actual exchange of any particle). 
Let us first consider the S wave contribution. The amplitude for the 
$V'\to V+$ scalar can be written as:

\bq
{\cal M}_0 = \eta'^\mu\eta^\nu t_{\mu\nu}. \nonumber
\eq

\noi Covariance allow us to write $t_{\mu\nu}$ in terms of $g_{\mu\nu}$ 
and the independent ($p',p$) four-momenta. Imposing $p'\cdot\eta'=0$, $p\cdot
\eta=0$ and using $p'^\mu t_{\mu\nu}=0$, which follows from the fact that 
$V'(p')$ is produced through a virtual photon in an $e^+e^-$ machine, we 
obtain:

\bq \label{eq:s0}
{\cal M}_0 = a_0 \left( \eta\cdot\eta' -{(p'\cdot\eta)(p\cdot\eta') \over 
p\cdot p'} \right ) 
\eq

Already at this point we encounter differences with the parametrizations used 
in the literature, where only the $\eta\cdot\eta'$ term is considered. Although
sizable effects are not produced by the extra term it is important to work 
with the proper Lorentz invariant amplitude. In particular, differences could 
become relevant when polarization measurements are involved. \footnote{Note 
that $a^{-2}_0\sum_{pol} | M_0 |^2 = 2 + \frac{m^2}{E^2}$, whereas $\sum_{pol}
(\eta \cdot \eta') = 2+ \frac{E^2}{m^2}$, with $m$ and $E$ the outgoing vector
meson mass and energy respectively.} 
\noi For the D wave contribution we write:

\bq \label{eq:m2}
{\cal M}_2 = A^{\mu\nu}\Pi_{\mu\nu\rho\sigma}B^{\rho\sigma}
\eq

%\bi
\noi where $A^{\mu\nu}$ describes the $V'(p')\to V(p)+D(Q)$, D standing for a  
spin two meson like object, $B^{\rho\sigma}$ describes the $D\to\pi\pi$ 
amplitude and $\Pi_{\mu\nu\rho\sigma}$ is the spin two projector:

\bq \label{eq:pola}
 \Pi_{\mu\nu\rho\sigma} \equiv \sum\limits_{\lambda=1}^5 h_{\mu\nu} 
 (\lambda)
 h_{\rho\sigma}(\lambda).  
\eq

\noi The polarization tensor $h_{\mu\nu}$ has the following properties:

\bq \label{eq:polaproper}
 h_{\mu\nu}(\lambda)  =  h_{\nu\mu}(\lambda),~~~~~~~~~~ Q^{\mu}h_{\mu\nu}
 (\lambda) ~ = ~ g^{\mu\nu} h_{\mu\nu}(\lambda)  =  0 
\eq

\bi
\noi Using these relations together with the projector property of 
$ \Pi_{\mu\nu\rho\sigma}$ one finds:

\bq \label{eq:projector}
 \Pi_{\mu\nu\rho\sigma}\equiv  \frac{1}{2} P_{\mu\rho} 
 P_{\nu\sigma} + \frac{1}{2} P_{\mu\sigma} P_{\nu\rho} - \frac{1}{3} 
 P_{\mu\nu} P_{\rho\sigma},  
\eq

\noi with 

\bq \label{eq:p}
 P_{\mu\nu} = g_{\mu\nu} - \frac{Q_{\mu}Q_{\nu}}{Q^2}.  
\eq

\noi On the other hand, by Lorentz covariance $B_{\rho\sigma}\propto q_\rho 
q_\sigma$. It is convenient to introduce:

\bq \label{eq:pimunu}
 \Pi_{\mu\nu} \equiv \Pi_{\mu\nu\rho\sigma} B^{\rho\sigma}
 = q_\mu q_\nu - \frac{1}{3}P_{\mu\nu} q^2.   
\eq

Furthermore, using Lorentz covariance and imposing again the conditions
 $p'\cdot\eta'=0$, $p\cdot\eta=0$ and $p'^\mu A_{\mu\nu}=0$, due to the 
coupling of $V'$ with a virtual photon, we find the  
general structure for $A_{\mu\nu}$, and through Eq. (\ref{eq:m2}) the di-pion 
D wave amplitude (Note that $Q^\mu \Pi_{\mu\nu}=Q^\nu \Pi_{\mu\nu}=0,$
therefore $p^{\prime\mu} \Pi_{\mu\nu}=p^\mu \Pi_{\mu\nu}$).

\bq  
 {\cal M}_2 = b \left(\eta'^\mu \eta^\mu - \frac{\eta^\nu 
 p^\mu}{p\cdot p'}(\eta' \cdot p)\right)\Pi_{\mu\nu} 
 + c \left( \eta'^\mu (\eta \cdot p') - p'^\mu 
\frac{(\eta \cdot p')(\eta' \cdot p)}{p\cdot p'} \right )
 p'^\nu \Pi_{\mu\nu} \nonumber
\eq

\bq \label{eq:am2} 
 \qquad \qquad + {a_2 \over m'^2 - m^2} \left((\eta\cdot \eta') - \frac{(\eta 
 \cdot p^\prime)(\eta^\prime \cdot p)}{p \cdot p^\prime}\right ) 
 p^{\prime\mu} p^{\prime\nu} \Pi_{\mu\nu}. 
\eq

\noi Where we introduced the $m'^2 - m^2$ factor to work with a dimensionless 
$a_2$.

\bi

\noi In order to obtain the decay rate, we carry out the integration 
over the $p_1, p_2$ Lorentz invariant phase space in the two pion center of 
mass reference frame, {\it i.e.} $\vec Q=0,~q_0=0$. We obtain:

\[ \quad \Gamma_0 \equiv \frac{1}{2 m'} \frac{1}{3} \int 
   \sum_{pol} |{\cal M}_0 \mid^{2} \frac{d^3 p}{{(2\pi)}^3 2p_0} \frac{d^3p_1}
   {{(2\pi)}^3 2p_{10}} \frac{d^3 p_2}{(2\pi)^3 2p_{20}}
 {(2\pi)}^4 \delta^4 (Q - p_1 - p_2) 
\]

\[ = \frac{1}{48\pi m'}  \int \sum_{pol} 
  |{\cal M}_0 \mid^{2} \frac{d^3 p}{{(2\pi)}^3 2p_0} (1 - 
   {\frac{4m^2_\pi}{Q^2})}^{\frac{1}{2}}. \]

The S-D wave interference vanishes upon integration. Indeed covariance imply

\[ \quad \Gamma_{int} \sim \frac{1}{3m'} \int 
  \sum_{pol} ( R_e {{\cal M}_0} A^{\mu\nu} \Pi_{\mu\nu}) 
 \frac{d^3p_1}{(2\pi)^3 2p_{10}} \frac{d^3 p_2}{(2\pi)^3 2p_{20}}  
  (2\pi)^4 \delta^4 (Q - p_1 - _2) \]

\[= ag_{\mu\nu}+bQ_\mu Q_\nu \]

\noi but using $Q^\mu \Pi_{\mu\nu}=g^{\mu\nu}\Pi_{\mu\nu}=0$, it follows that
$a=b=0$. 

\bi
\noi For the $D$ wave we proceed along the same lines.

\[ \Gamma_2  = \frac{1}{2m'} \frac{1}{3} \sum_{pol} \int 
  \frac{d^3 p}{{(2\pi})^3 (2p_0)} \big( A^{\mu\nu} {A^*}^{\rho\sigma} \big)
   \overline{\Pi}_{\mu\nu\rho\sigma} 
\]

\noi where 

\[ \quad  \overline{\Pi}_{\mu\nu\rho\sigma} = \int \sum_{pol}
(\Pi_{\mu\nu} \Pi_{\rho\sigma}) 
 \frac{d^3 p_1}{{(2\pi)}^3 (2p_{10})} 
 \frac{d^3 p_2}{{(2\pi)}^3 (2p_{20})} {(2\pi)}^4 \delta^4 (Q-p_1-p_2)
\]

\[={\bf x}\Pi_{\mu\nu\rho\sigma}.\] {\bf x} is determined using the reference 
frame where  $\vec Q=0,~q_0=0$. For example

\[  \overline{\Pi}_{3333} =\frac{2}{3} {\bf x}=\int \sum_{pol}
(\Pi_{33} \Pi_{33}) 
 \frac{d^3 p_1}{{(2\pi)}^3 (2p_{10})} 
 \frac{d^3 p_2}{{(2\pi)}^3 (2p_{20})} {(2\pi)}^4 \delta^4 (Q-p_1-p_2),
\]

\noi in this way we obtain

\[ 
 {\bf x}=\frac {Q^4}{60\pi} ( 1-\frac{4m^2_{\pi}}{Q^2})^{\frac{5}{2}} 
\]

We observe that using the $q_0=0, \vec{Q} =0$ reference frame, it is easy to 
show that $\Pi_{\mu\nu} \sim Y_2(\theta,\phi)$, {\it i.e.} it is associated to
the di-pion $D$ wave. This together with the fact that $a_2, b$ and $c$ can 
only depend upon $s\equiv (p_1+p_2)^2$, allow us to conclude that ${\cal M}_2$
describes the di-pion spin 2 wave. In the following we consider the particular
 case where only one Lorentz invariant amplitude is included in $A_{\mu\nu}$. 
To this end we set $b=c=0$ in Eq(\ref{eq:am2}). This is the simplest way to 
consistently introduce $D$ wave effects in the invariant amplitude. Within 
this approximation we obtain for the di-pion invariant mass distribution 
($s\equiv (p_1+p_2)^2 = Q^2$).

\bq \label{espectro}
{d\Gamma \over d\sqrt s}=\frac{(m'p)\sqrt{s}}{3(4m'\pi)^3}
\sum_{pol}|(\eta\cdot\eta') -{(\eta\cdot p')(\eta'\cdot p) \over p\cdot
 p' }|^2~(SW + DW)
\eq

\bi
\noi where $p$ stands for the three momentum ($p~=~|\vec{p}|$) and

\bq \label{sw}
 SW=|a_0(s)|^2(1-{4m^2_\pi\over s})^{1\over 2}
\eq

\bq\label{dw}
 DW= \frac{|{a_2(s) \over m'^2 - m^2}|^2}{180} (1-{4m^2_\pi\over s})^{5\over 2}
 \left [(s-(m'^2+m^2))^2-4m'^2m^2 \right ]^2 
\eq

\bi
\noi with 

\bq
 2m'p = \left [ (s-m'^2-m^2)^2-4m'^2 m^2 \right ]^{1\over 2}.
\eq

\bq
 \sum_{pol} |\eta \cdot \eta' - {(\eta\cdot p')(\eta'\cdot p) \over p\cdot
 p'}|^2 =2+ {4 m^2 m'^2 \over (m^2 + m'^2 - s)^2}
\eq

We have now a  general expression describing the decay $V'\to V\pi\pi$, which  
involves two invariant amplitudes, associated to the $S$ and $D$ wave 
respectively.  We assume, when $\sqrt{s} \leq 0.9~GeV$,  the $S$ wave is 
composed of the $f_0(600)$ and a non resonant background. This is
an approximation since other resonances could contribute to the amplitude in 
the kinematical region considered. However, it is reasonable to expect that 
the contribution of the $f_0(980)$  ($\Gamma_{f_0(980)}\approx 100$ MeV) and 
higher states decaying in two pions behave softly in the neighborhood of the 
$f_0(600)$, even if the latter is a broad resonance. Crossed channel 
contributions are treated in a similar way, {\it i.e.} considered as soft 
functions of the di-pion invariant mass. Therefore we parametrize the form 
factor associated to the $S$ wave in the following way.
\bi

\be
a_0^{(i)}=\left( {a^{(i)}m_0^2 \over D(s)}+ {b^{(i)}\over 1-{c^{(i)}s \over 
m_0^2 }} \right),\label{a0}
\en

\noi $m_0$ is introduced for dimensional reasons and is fixed to
$m_0=0.5~ GeV$. For $D(s)$ we used two different expressions:

\bq\label{propa}
D(s)&=&s-m^2+\Pi(s)  \nonumber  \\
D(s)&=&s-m_p^2+im_p\Gamma_p
\eq 

\noi where $\Pi(s)$ stands for the $f_0(600)$ self energy which involves loop
of kaons and pions. Note that Eq(\ref{propa}) defines the mass~$m_p$~and 
width~$\Gamma_p$~of
the resonance in terms of the real and imaginary part of the pole of the $S$ 
matrix. If a pole exist, it should be independent of the process 
where it is observed. However neither the residue nor the background have to 
be the same for different processes. For this reason we include the index $i$
 which is associated to the physical decay under consideration. In the 
kinematical region of interest the $D$ channel is non-resonant, thus we can 
approximate it by a soft background:

\be
a_2^{(i)}=\left( {f^{(i)}\over 1-{g^{(i)}s \over 
m_0^2 }} \right) \label{a2}
\en
It should be clear that our approach is a phenomenological, and that we have 
not attempted to explicitly incorporate the scale anomaly. In fact, it can 
be argued that such a contribution is included within the background. 
\bi

\section{The fit}.

We consider the following decays: $\Upsilon(3S) \to \Upsilon(1S)+
\pi\pi$ \cite{Butler}, $\Upsilon(3S) \to \Upsilon(2S)+\pi\pi$ \cite{Butler}, 
$\Upsilon(2S) \to \Upsilon(1S)+\pi\pi$ \cite{Butler,alexander}, $\psi(2S) \to 
J/\psi +\pi\pi$ \cite{bai}, resulting in a total of 165 points. All the data 
points but the last have been gotten from plots since no listings of the data 
are available.  It is
important to mention that previous fits \cite{Komada} of the $\psi(2S) \to 
J/\psi +\pi\pi$ included only a subset of the experimental data. 
\bi

Following the analysis in \cite{alexander,bai} we first attempted a fit in 
terms of
the scale anomaly \cite{Voloshin,Novikov}. The fit produces a $\chi_{d.o.f.}^2
~>~2$. This is not an unexpected result, the amplitude associated to the scale
anomaly is derived under the assumption that the pions are soft, and this is 
not the case for all the data under consideration. In fact, according to 
\cite{Komada} and previous analysis \cite{pdg}, the $f_0(600)$ is expected
to contribute to these processes. This is the motivation to try a fit in terms
of the pole approach, as discussed in the previous section. 
\bi

The fit using the pole approach is based on Eqs(\ref{espectro},\ref{sw},
\ref{dw},\ref{a0},\ref{propa},\ref{a2}). The pole can be parametrized in terms 
of the full propagator, including the self-energy $\Pi(s)$, which involves
loops of kaons and pions. In such case, as a result of the fit, we get a 
mass and the $g_{fKK}$ and $g_{f\pi\pi}$ coupling constants, and from these 
the pole position is determined. Unfortunately, the fit turns out to be 
insensitive to the values of the coupling constants, {\it i.e.} different 
values of the $g's$ lead to essentially the same fit, but correspond to pole 
positions far from each other. For this reason our analysis is restricted 
to $D(s)$ given by Eq(\ref{propa}) where the mass ($m_p$)and width 
($\Gamma_p$)are obtained directly from the fit.
\bi

The fit involves 33 parameters. We consider four processes ($i=1$ to $4$
in Eqs(\ref{a0}, \ref{a2})) and for each of these we require seven parameters 
($b$ and $f$ which are complex, and $a,c,g$). Three normalization factors are 
also parameters of the fit -because some of the reported data refers to number
 of events, not to a differential decay rate- and finally the mass and width 
of the resonance. The fit leads to a pole located at $m_p=528 \pm 32$ MeV and 
$\Gamma_p=413 \pm 45$ MeV together with the parameters reported in Table(1), 
and a $\chi_{d.o.f.}^2=1.12$. The result of the joint fit are shown as the 
solid lines in Fig(1). We do not report the result of the fit for S wave 
alone, however we should mention that including D wave effects improves the 
total $\chi^2$ but the $\chi_{d.o.f.}^2$ remains unchanged. Note in this 
respect that the fit parameters are obtained from  the di-pion invariant 
mass spectrum, the angular distributions are not involved. Once the fit 
parameters are fixed, numerical integration over the di-pion invariant mass is
performed and we obtain -up to normalization again- the following angular 
distributions (in GeV units) and the curves shown in Fig(3):

\be
\Upsilon(2S) \to \Upsilon(1S)\pi\pi\cite{alexander}:~~~{d\Gamma \over d(cos\theta)}=0.07(0.0674(3cos^2\theta~-1)^2+1.09) \label{ad2s}
\en

\be
\Upsilon(3S) \to \Upsilon(1S)\pi\pi\cite{Butler}:~~~{d\Gamma \over d(cos\theta)}=2.2(0.0009(3cos^2\theta~-1)^2+0.511)  \label{ad3s}
\en

\be
\Psi(2S) \to J/\Psi\pi\pi\cite{bai}:~~~{d\Gamma \over d(cos\theta)}=6.5(5.49(3cos^2\theta~-1)^2+359)  \label{adpsi}
\en

\bi

Besides the pole position and the quality measure through the $\chi_{d.o.f.}^
2$, it is instructive to analyze the different contributions 
to the decay rate. These are  shown in Fig(2) where the dots corresponds
to the pole contribution, the dashed line to the background and resulting fit
is represented by the continuous line. From these plots 
one can see that the pole and background contributions must interfere
destructively in all cases, except the $\Upsilon(3S) \to \Upsilon(1S)$
transition for which the kinematical region includes the pole position and 
both, constructive (below the pole-mass) and destructive (above the pole) 
interferences occur leading thus to the structure around 650 MeV. In table(2)
we quote the contributions to the $\chi^2$ from the different processes.

\section{Summary}.

In this paper we consider the phenomenological description of the
following decays:
$\Upsilon(3S) \to \Upsilon(1S)+\pi+\pi$, $\Upsilon(3S) \to \Upsilon(2S)+\pi+
\pi$, $\Upsilon(2S) \to \Upsilon(1S)+\pi+\pi$, $\psi(2S) \to J/\psi +\pi+\pi$.
As far as we can see, it is not possible to obtain a good quality fit in terms
of the scale anomaly alone. Using general arguments, we derived an expression 
for the invariant amplitude describing flavor conserving processes of the 
type $V' \to V+\pi+\pi$, including both S and D waves, which involves two
invariant amplitudes. We parametrized the S wave form factor with a pole plus 
a soft background and the D wave form factor by pure soft background. 
Fitting the 
data yields a pole in $m_p=528 \pm 32 MeV$ and $\Gamma_p=413 \pm 45 MeV$ with  
$\chi_{d.o.f.}^2=1.12$. We remark that a strong interference among the pole and
the background is required to fit the data. Thus, our analysis seems to
indicate that physics in S wave pion-pion interaction below 800 MeV is 
governed by a subtle interplay between $f_0(600)$ meson contributions and
a big background, difficult to understand in terms of conventional physics
and which could be associated to the scale anomaly.  

\bi

\section{Acknowledgments}
This work is supported, in part, by CONACyT under grant 37234-E. One of the 
authors (J.L.L.M.) acknowledges support from the ICTP, Trieste where part of 
this work was done.

\newpage

\begin{center}
FIGURE CAPTION.
\end{center}
\bi

\noi Figure 1. Data points used in the analysis 
\cite{Butler,alexander,bai} and the resulting fit (solid line). In the 
horizontal axes we plot $m_{\pi\pi}=\sqrt{s}$ whereas the vertical refers 
to the 
differential decay rate ${d\Gamma \over d\sqrt{s}}$, or number of events,
as shown in the plots and discussed in the quoted references. For the same 
reaction, open circles and solid triangles refer to data obtained from 
exclusive and inclusive processes respectively. 
\bi

\noi Figure 2. The curve resulting from the fit to the data
(solid line), and for comparison the pole (dashed line) and background
(dotted line) contributions. Note that for the $\Upsilon(3S) \to \Upsilon(2S)+\pi\pi$ the fit-curve is close to the axes, which implies a strong destructive
interference among the pole and the background contributions.
\bi

\noi Figure 3. Angular distribution as obtained from 
Eqs(\ref{ad2s},\ref{ad3s},\ref{adpsi}) and compared to data from
Ref(\cite{alexander,Butler,bai}). Open circles and solid triangles refer
to data obtained from exclusive and inclusive processes respectively. 
\bi\bi

\begin{center}
TABLE CAPTION.
\end{center}
\bi

\noi Table 1. Parameters resulting from the fit. The 
normalization factors refer to: $N_a$  data from Ref(\cite{alexander}), 
$N_2$ to the $\Upsilon(3S)\to \Upsilon(1S)+\pi^0\pi^0$ and $N_3$ to
$\Upsilon(3S)\to \Upsilon(2S)+\pi^0\pi^0$. The parameters are defined
by Eqs(\ref{a0}, \ref{a2}). Processes are labeled 1:
$\Upsilon(2S)\to\Upsilon(1S)\pi\pi$, 2: $\Upsilon(3S)\to\Upsilon(1S)\pi\pi$
3: $\Upsilon(3S)\to\Upsilon(2S)\pi\pi$ and 4: $\Psi(2S)\to J/\Psi\pi\pi$.
\bi

\noi Table 2. $\chi_{d.o.f.}^2$ for each separate process as obtained from the 
fit to the di-pion invariant mass distribution.

%\newpage
%\pagenumbering{alph}
%\setcounter{page}{28}
\begin{figure}[btp]
\centerline{\epsfxsize = 475 pt
\epsfbox{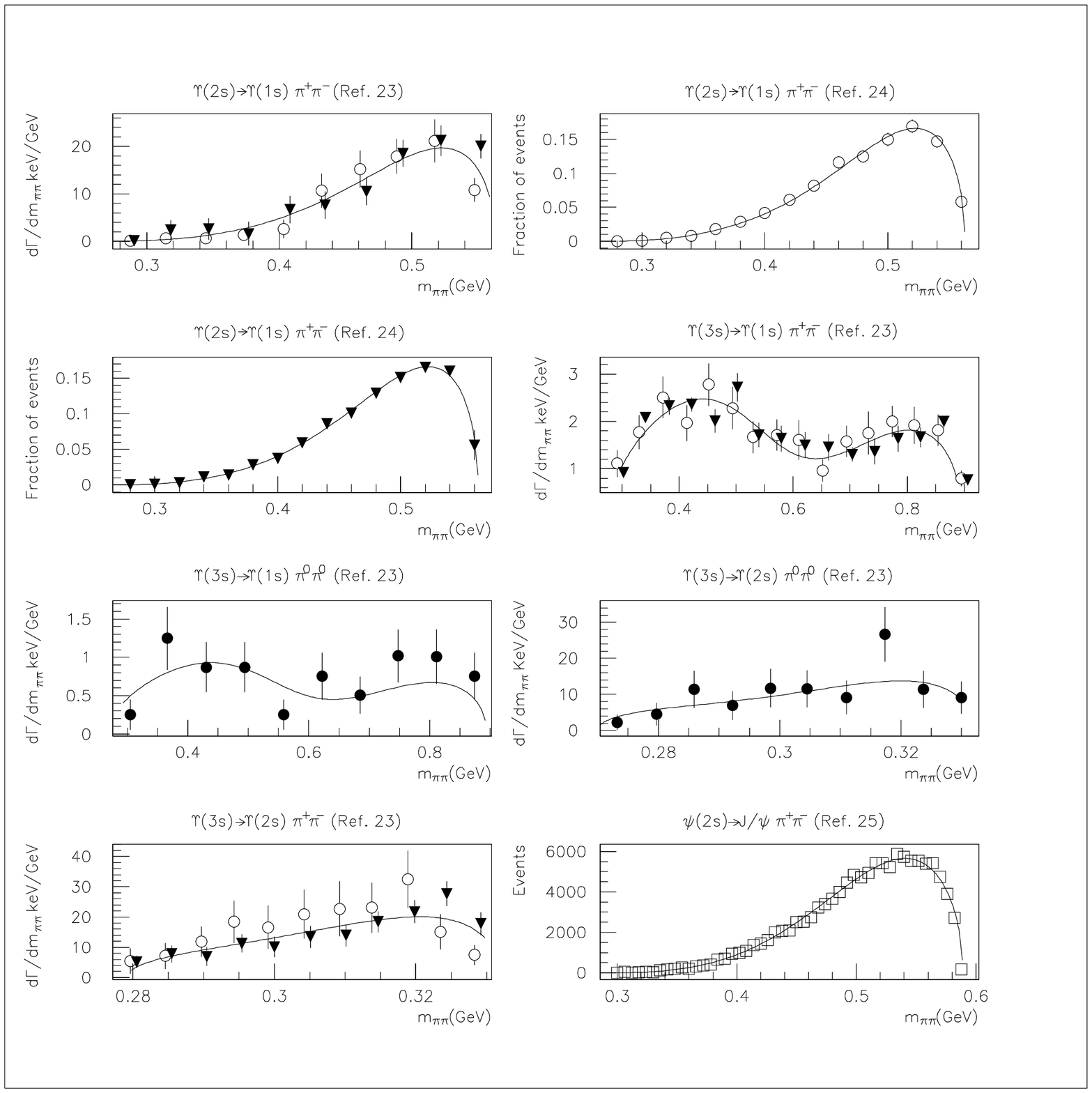}}
%\vspace{0cm}
\begin{center}
\noi {\footnotesize\bf {Figure 1}}
\end{center}
\end{figure}

%\newpage[btp]
\begin{figure}
\centerline{\epsfxsize = 475 pt
\epsfbox{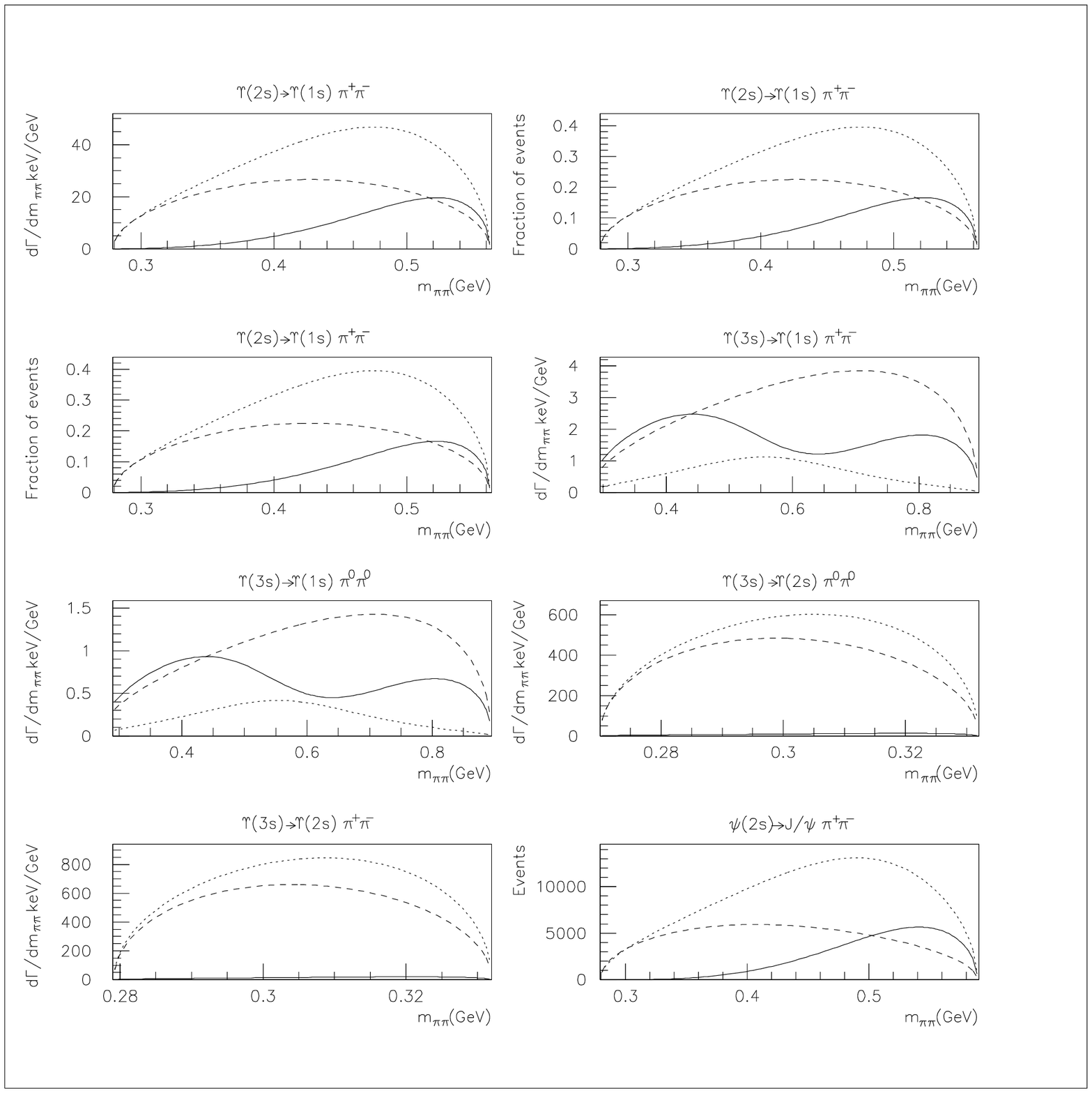}}
%\vspace{0cm}
\begin{center}
\noi {\footnotesize\bf {Figure 2.}}
\end{center}
\end{figure}

\begin{figure}
\centerline{\epsfxsize = 350 pt
\epsfbox{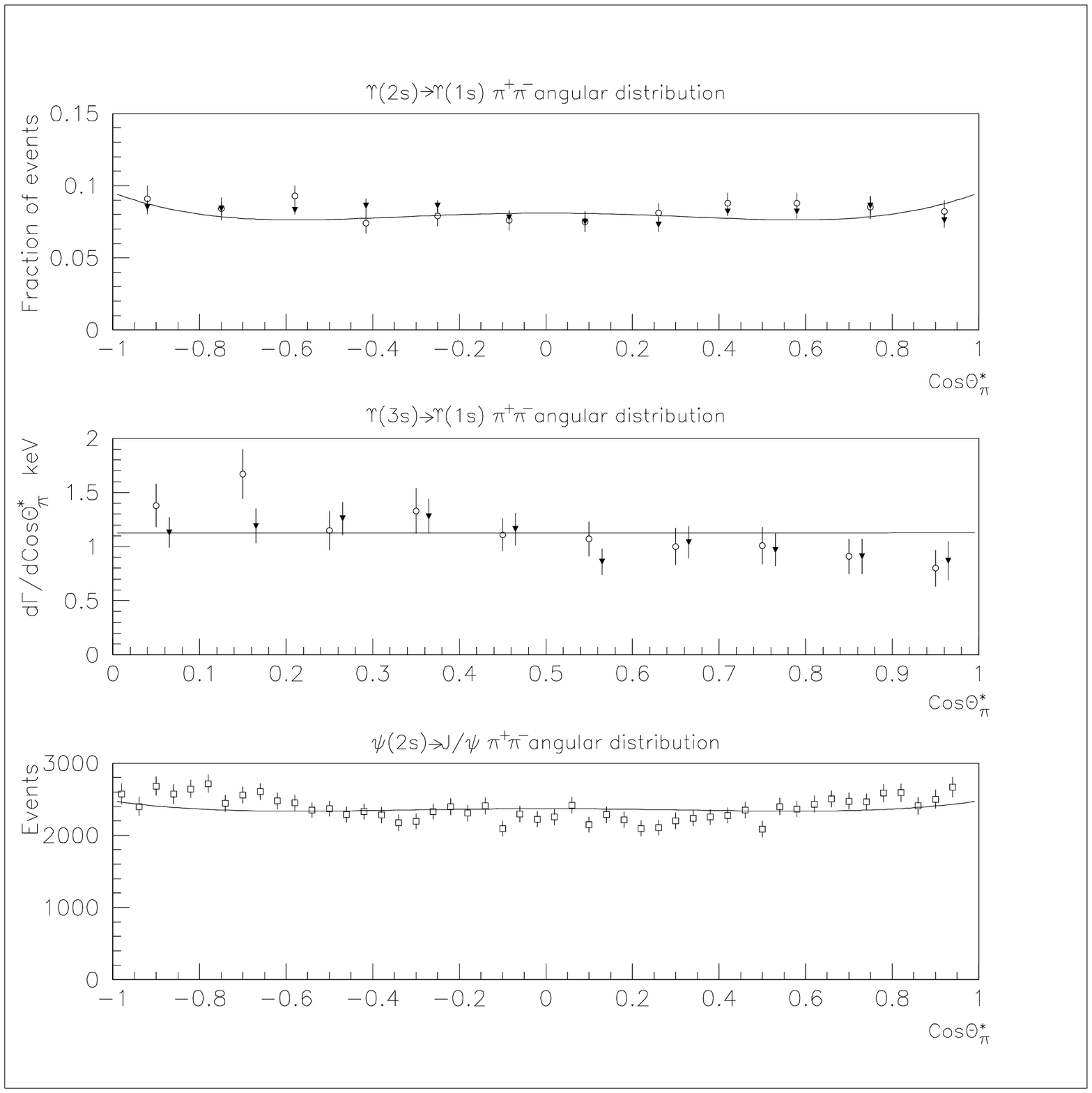}}
%\vspace{0cm}
\begin{center}
\noi {\footnotesize\bf {Figure 3.}}
\end{center}
\end{figure}

\newpage
\qquad \qquad \qquad \qquad 
\begin{tabular}
[c]{|c|c|}\hline
$N_{a}$ & $8.4$x$10^{-3}$\\\hline
$N_{2}$ & $0.36$\\\hline
$N_{3}$ & $0.62$\\\hline
$a_{1}$ & $8.2$x$10^{3}$\\\hline
$b_{1}$ & $5.3$x$10^{3}+i5.1$x$10^{3}$\\\hline
$c_{1}$ & $-0.14$\\\hline
$f_{1}$ & $-1.6$x$10^{3}+i3.8$x$10^{3}$\\\hline
$g_{1}$ & $0.4$\\\hline
$a_{2}$ & $7.6$x$10^{2}$\\\hline
$b_{2}$ & $-1.1$x$10^{3}+i8.4$x$10^{2}$\\\hline
$c_{2}$ & $5.4$x$10^{-2}$\\\hline
$f_{2}$ & $2.3+i2.8$\\\hline
$g_{2}$ & $0.33$\\\hline
$a_{3}$ & $-1.2$x$10^{5}$\\\hline
$b_{3}$ & $-7.8$x$10^{4}-i1.0$x$10^{5}$\\\hline
$c_{3}$ & $-1.1$\\\hline
$f_{3}$ & $-4.8$x$10^{3}-i9.0$x$10^{5}$\\\hline
$g_{3}$ & $-5.0$x$10^{4}$\\\hline
$a_{4}$ & $4.8$x$10^{4}$\\\hline
$b_{4}$ & $3.2$x$10^{4}+i3.7$x$10^{4}$\\\hline
$c_{4}$ & $-0.48$\\\hline
$f_{4}$ & $3.2$x$10^{4}-i3.2$x$10^{4}$\\\hline
$g_{4}$ & $-4.9$x$10^{-2}$\\\hline
\end{tabular}     
\begin{center}
\noi{\footnotesize {Table 1.}}
\end{center}
\bi

\begin{center}
\begin{tabular}
[c]{|c|c|c|}\hline
Process & $N_{data}$ & $\chi_{d.o.f.}^{2}$\\\hline
$\Upsilon(2s)\rightarrow\Upsilon(1s)\pi\pi$ & 48 & 0.74\\\hline
$\Upsilon(3s)\rightarrow\Upsilon(1s)\pi\pi$ & 40 & 1.3\\\hline
$\Upsilon(3s)\rightarrow\Upsilon(2s)\pi\pi$ & 32 & 1.1\\\hline
$\psi(2s)\rightarrow J/\psi\pi\pi$ & 45 & 1.5\\\hline
\end{tabular}
\end{center}
\begin{center}
\noi{\footnotesize {Table 2.}}
\end{center}

\end{document}